\documentclass[preprint]{aastex}

\usepackage{natbib}

\shorttitle{C$_{3}$ in 9P/Tempel 1 during {\em DI}}
\shortauthors{\'Ad\'amkovics et al.}

\begin{document}

\title{The changing rotational excitation of C$_3$ in comet 9P/Tempel 1  during {\em Deep Impact}}

\author{M\'at\'e \'Ad\'amkovics, Imke de Pater, and Hy Spinrad}
\affil{Astronomy Department, University of California,
    Berkeley, CA 94720}
\email{mate@berkeley.edu}

\begin{abstract}
The 4050\AA\ band of C$_3$ was observed with Keck/HIRES echelle spectrometer during the {\em Deep Impact} encounter. We perform a 2-dimensional analysis of the exposures in order to study the spatial, spectral, and temporal changes in the emission spectrum of C$_3$. The rotational population distribution changes after impact, beginning with an excitation temperature of $\sim$45\,K at impact and increasing for 2\,hr up to a maximum of 61$\pm$5\,K.  From 2 to 4 hours after impact, the excitation temperature decreases to the pre-impact value. We measured the quiescent production rate of C$_3$ before the encounter to be 1.0~$\times~10^{23}$~s$^{-1}$, while 2 hours after impact we recorded a peak production rate of 1.7~$\times~10^{23}$~s$^{-1}$.  Whereas the excitation temperature returned to the pre-impact value during the observations, the production rate remained elevated, decreasing slowly, until the end of the 4\,hr observations. These results are interpreted in terms of changing gas densities in the coma and short-term changes in the primary chemical production mechanism for C$_{3}$. 
\end{abstract}


\section{Introduction}

The {\em Deep Impact (DI)} mission created an artificial outburst on comet 9P/Tempel 1 by excavating material from the subsurface \citep{Ahearn2005}. The event was observed by a network of orbiting (e.g., \citet{Feldman2006, Lisse2006}) and ground-based facilities \citep{Meech2005}. 
 \citet{Feldman2006} use ultraviolet observations to show that the amount of CO relative to H$_2$O is similar in both the impact ejecta and in the outgassing of the quiescent comet.
 \citet{Jehin2006} measure $^{12}$C/$^{13}$C and $^{14}$N/$^{15}$N in the material released by impact and find it to be the same as the surface material. 
 \citet{Mumma2005} show that the relative abundances of CH$_{3}$OH and HCN are the same before and after impact, although the amount of C$_2$H$_6$ relative to H$_2$O increases by almost a factor of two shortly after impact. 
 \citet{Schleicher2006} report that in nearly all respects --- including the production rate of C$_3$ --- comet 9P/Tempel 1 returned to pre-impact conditions within 6 days. 
 With the precisely timed outburst caused by {\em DI}, the short (4\,hr) time period directly after impact offers a unique opportunity to  observe molecular formation mechanisms that occur on short timescales. Indeed,  \citet{Jackson2009} use two- or three-step Haser models to study temporal changes in the total emission from the following species: O, OH, CN, C$_2$, C$_3$, NH, and NH$_2$. Here we describe the observed variations in the excitation temperature and production rate of C$_3$ in the first four hours after {\em DI}. 
 
\section{Observations and Reduction}

The observations were conducted using the HIRES cross-dispersed echelle spectrometer \citep{Vogt1994} at the W.M. Keck observatory.  HIRES offers nearly complete spectral coverage from 3000 -- 5880\,\AA\ at a spectral resolution, $\lambda / \Delta \lambda \sim$ 48,000 with a 7$\farcs$0 $\times$ 0$\farcs$86 slit. The plate scale is 0$\farcs$239. Three 20\,min exposures were taken on 30 May 2005 UT and combined to characterize the comet before impact. On 04 Jul 2005 UT, a series of shorter exposures were taken to capture short timescale changes in the spectrum after impact. Exposure times increase with airmass and range from 10 -- 30 min during the first 3 hrs after impact. An analysis and description of the observations is given in \citet{Jehin2006}, with additional details in \citet{Cochran2007} and \citet{Jackson2009}. Here we focus on the 4050\,\AA\ band of C$_3$ (from 4049 -- 4057\,\AA), which falls onto the central 1/8 of an echelle order, near the peak of aperture blaze. 

Standard procedures are used to bias correct and flat field each target and calibration exposure. Cosmic rays are removed using a two-step process. First, large cosmic ray hits are removed by considering the 13 target exposures together at each pixel and replacing large deviations from the mean with the median value. Then, to identify smaller cosmic ray hits, a 6$\times$6 pixel spatial filter is used on each image, where pixels exceeding the mean value within the box by 2.5$\sigma$ are replaced by the mean value. The echelle order is linear along the CCD over this wavelength range and it is sufficient to rectify the spectra by rotating the image 1.15$^{\circ}$. Rectification is verified to be better than a pixel by comparing the dust continuum spatial profiles along the order. Flux calibration is performed for both the pre-impact and post-impact exposures using exposures of the flux standards, Feige 67 and BD+284211, respectively \citep{Oke1990}. The total counts are integrated along the slit for the calibration standard and averaged over the bandpass of the \citet{Oke1990} observations (4020-4060\,\AA). The HIRES observations confirm that there are no significant spectral features in either of the calibration stars for this bandpass, so the mean observed count rates and the literature values for the total flux at 4050\,\AA\ are used to convert from observed count rate (DN/s) to flux density (ergs s$^{-1}$ cm$^{-2}$ Hz$^{-1}$). We correct for the different airmass during the target and calibration exposures using a characteristic atmospheric extinction, $k=0.56$, at 4000\,\AA.

The solar continuum dominates the cometary spectrum at most of the sampled wavelengths, with temporal variations in continuum reflectivity due to the changing particle size and composition of the ejected material. We use the median value along the dispersion axis to calculate the spatial profile of reflected sunlight at each time step to empirically remove the solar contribution.  The median profile is convolved with a high-resolution, $R\sim85,000$, UVES solar spectrum\footnote[2]{http://www.eso.org/observing/dfo/quality/UVES/pipeline/solar\_spectrum.html} and interpolated onto the HIRES plate scale to model the reflected sunlight. In the narrow wavelength range being considered, the changes in the reflectivity spectrum of the ejecta are small and all observations are successfully fit with one spectral template. This model is subtracted from the observations, leaving only the gas emission. The higher signal-to-noise spectra taken on 30 May 2005 UT are used to demonstrate the dust continuum removal, Figure~1. Before removing the solar spectrum, some transitions in the emission spectrum of C$_{3}$ (e.g., near 4052.5 and 4053.2\,\AA) are not apparent in the observations because they are collocated with solar features. After removing the solar spectrum these lines stand out against the background.

\section{Temperatures and Production Rates}

The C$_{3}$ excitation profile --- that is, the distribution of population among rotational levels ---  can depend on the temperature and density in the coma, the heliocentric velocity and distance, and the formation mechanism of the molecule (e.g., the parent and grandparent molecules of C$_{3}$). \citet{Rouss2001} describe a statistical equilibrium model that can be used to interpret the rotationally-resolved emission spectrum of cometary C$_{3}$ by taking into account changes in heliocentric velocities and distances. However, changes in those values are small on the timescale of these observations and we assume a simple thermal distribution of states to model the emission spectrum. This approach is sufficient for the purposes of measuring the relative changes in the rotational excitation profile among exposures. 

Spectra of C$_3$ are calculated using the line list of \citet{Tanabashi2005} for the $A ^1\Pi_u \leftarrow X^1\Sigma^+_g$ (000-000) transitions. The fraction of the total molecules that are in a particular $J$ level is given by
\begin{equation}
F_J(T_r) = \frac{2J+1}{q_r}exp\left(\frac{-hcB_0}{kT_r}J(J+1)\right)
\end{equation}
where $q_r$ is the rotational partition function given by
\begin{equation}
q_r = \sum_{J\: even}^{\infty}(2J+1)e^{-hcB_0J(J+1)/kT_r}
\end{equation}
For C$_3$ the ground state rotational constant $B_0$=0.43057  cm$^{-1}$ \citep{Schmutt1990}. 

Each modeled spectrum (e.g., Figure~2) has three free parameters; the excitation temperature, $T_{r}$, line width, $w$, and an intensity scaling factor, $\alpha$, which are used to reproduce the observed spectra (Table 1). Spectra are calculated for a range of $T_{r}$, $w$, and $\alpha$ and compared to observations. The quality of fit is determined by $\chi^{2} = \sum_{i} (\delta y^{2}_{i}/\sigma^{2})$, where $\delta y_{i}$ is the residual at each sampled wavelength, $i$, $\sigma$ is approximated by the noise in the observed spectrum, and the best fit corresponds to $\chi^2_{min}$. Locations in parameter space where $\chi^2 - \chi^2_{min} = 1$ give the uncertainties in the input parameters. The residuals from the fit indicate that the thermal population distribution is moderately successful at reproducing the observations. A successful fit would be indicated by residuals that are random noise, however there is structure in the residuals  due to the non-thermal population distribution. 

We integrate the flux from C$_3$ over 4049 -- 4058\,\AA\ and use the standard fluorescence efficiency factor of $g(r_{H}) = 1.0\times10^{-12}\times r_H^{-2}$, at a heliocentric distance, $r_H =$ 1.55\,AU \citep{Ahearn1995}, to find the total number of C$_3$ molecules in the sampled area. A Haser model \citep{Haser1957} assumes isotropic outgassing from the nucleus at a constant velocity, $v$. Parent species are destroyed to form radicals with lifetimes of $\tau_p$ and $\tau_r$, which define characteristic length scales $l_p = \tau_p v$  and $l_r = \tau_r v$ for the parent and radical, respectively. The parent species and radical are assumed to travel in the same direction and at the same speed. The number density of species at a distance, $x$, from the nucleus is given by
\begin{equation}
n(x) =  \frac{Q}{4\pi v x^2} \left( \frac{l_r}{l_p - l_r} \right) \left(  e^{x/l_p}  - e^{x/l_r} \right)
\end{equation}
As detailed in \citet{Newburn1984} the relationship between the production rate, $Q$~(s$^{-1}$) and the column density profile, $N(x)$ (cm$^{-2}$), is given by
\begin{equation}
N(x) = \frac{Q}{4\pi v} \frac{2}{x} \left( \frac{l_r}{l_p - l_r} \right) \times 
\left[ \int^{x/l_r}_{0} k_0(y)dy  - \int^{x/l_p}_{0} k_0(y)dy \right]
\end{equation}
where $x$ is the distance from nucleus (cm), and $k_0$ is the zero-order modified Bessel function of the second kind.

 The projected area that we observe is small, so these observations are not useful for constraining scale lengths and we use the values of $l_p = 6.7 \times 10^{3}$\,km and $l_r = 6.5 \times 10^{4}$\,km at a $r_H$=1.55\,AU \citep{Ahearn1995}. 

The spectra taken were recorded in 13 exposures over the first 4\,hrs after impact (until the comet set), and when extracted along the slit show temporal changes in both excitation profile (Figure~3A), and the total flux  \citep{Cochran2007,Jackson2009}. The high-J lines in the $R$-branch band-head (near 4050\,\AA) are more prominent in the exposures roughly 100\,min after impact (Figure~3A). The spatial profiles along the slit also change shortly after impact (Figure~3B). Three examples of the radial dependence of the C$_3$ column density are shown in Figures 3C -- 3E, along with the best fit production rates, $Q$, and corresponding model profiles. The C$_{3}$ production rates determined using this technique fall in the range of 1.1 -- 1.7$\times 10^{23}$\,s$^{-1}$, which are consistent with the pre-impact measurements of \citet{Lara2006}, however they are an order of magnitude smaller than the imaging measurements of \citet{Schleicher2006} and the analysis of \citet{Cochran2009}. Discrepancies are primarily due to the choice of scale length, $l_r$,  which is a factor of 5.5 smaller than used by  \citet{Cochran2009}. The temporal changes in the C$_3$ production rates directly after impact are shown in Figure 4 (bottom).

The trend in production rates is generally consistent with integrated flux of C$_3$ emission presented in Figure~5 of \citet{JacksonCochran2009}. In both datasets there are no changes over pre-impact conditions for the first 50min (3000s) of observations before a monotonic increase in production rates (or integrated flux) over $\sim$130min (7800s). The peak in production rate and integrated flux are a factor of $\sim$2.5 larger than the pre-impact values. After the peak there is a very gradual decrease until the end of the observations. 

\section{Discussion \& Conclusions}

Since the heliocentric velocity of Tempel 1 does not change significantly over the 4\,hrs of the observations, the excitation spectrum does not change due to the Swings effect (e.g., changes in the radiative excitation of C$_3$ caused by a Doppler shift of absorption lines in the solar spectrum) and some other process must account for the observed change in the excitation temperature, $T_{r}$, after impact (Figure~4). Variations in the gas density around the ejecta are one way to explain changes in the excitation profile. C$_3$ lacks a permanent dipole moment so radiative relaxation is not an efficient method for removing population from high-$J$ states. At high densities however, collisional de-excitation can thermalize the population distribution. \citet{ABM2003} have shown that in the interstellar medium the C$_3$ excitation profile depends on density, such that $T_r$ exceeds the kinetic temperature at densities below 500\,cm$^{-3}$. 

If the gas density in the coma directly after impact is large, then $T_r$ is essentially indicative of the thermal temperature. We measured $T_r\sim$45\,K right before impact,  which then increased up to 60\,K at 100\,min after impact --- the same temperature it was in May. This increase may be due to decreasing gas density and the lack of collisional de-excitation. However, there is then a puzzling decrease in $T_{r}$ from 100 -- 200\,min after impact. If the gas density were monotonically decreasing, then T$_r$ should increase and then plateau.  However, there is the possibility that the 15\,K change in $T_r$ occurs independently of the changes caused by {\em DI}. The quiescent $T_r$ measured on 30 May 2005 is the same as the peak post-impact value of 60\,K, so that perhaps  $T_r$ varies with time for C$_{3}$ . The decrease in $T_r$ at times greater than 100\,min after impact supports the possibility that $T_{r}$ is gradually fluctuating on a timescale of hours.

The production rate for C$_3$ reaches a plateau 130\,min after {\em DI} and then appears to decline after $\sim$183\,min. The relative flux of CN follows a similar progression with time \citep{Jehin2006}, however CN reaches maximum production at a mid-exposure time of 96\,min, significantly earlier than C$_3$. One simplistic interpretation is that there are additional intermediate reactions between the photodissociation of the parent molecule and the formation of C$_{3}$. The primary production pathway of cometary C$_3$ is the photodissociation of either propyne (H$_3$CCCH) or allene (H$_2$CCCH$_2$) --- in either case, one of the isomers of C$_3$H$_4$ leads to the formation of C$_3$ via the C$_3$H$_2$ radical intermediate \citep{Helbert2005}. This common radical means that multiple pathways can produce the same rotational excitation spectrum of C$_3$ \citep{Song1994} and so the parent molecule of C$_3$ cannot be distinguished. \citet{Jackson2009} use a 3-step chemical model with the photodissociation of the C$_3$H$_2$ radical as the production pathway for C$_3$, and consider either allene or propyne as the precursor to the parent radical. However, other mechanisms have been hypothesized for the formation of C$_{3}$, which could produce C$_{3}$ with a different excitation spectrum. \citet{Helbert2005} mention the electron impact dissociation of C$_{3}$H$_{4}$ as a source of C$_{3}$ but note that rates for individual reactions have not been determined and hence the relevance of this mechanism is uncertain. The propynal radical (C$_{3}$H$_{2}$O) has also been proposed as a source of C$_{3}$ \citep{Kras1991}, and proceeds via an excited state intermediate, C$_{3}$H$_{2}^{*}$. This radical could be the parent molecule of C$_{3}$ with a different $T_{r}$ than when produced from C$_{3}$H$_{2}$. In general, the total yield of C$_{3}$ from C$_{3}$H$_{2}$O is only $\sim$1\% of the production from C$_{3}$H$_{2}$, yet this mechanism may have an increased relevance in the  1 -- 2\, hrs after {\em DI}. Similarly, the dissociative recombination of C$_{3}$H$_{5}^{+}$ may yield C$_{3}$ with a $T_{r}$ that is larger than when produced by the photodissociation of C$_{3}$H$_{2}$. It is unclear if there is link between the fact that $Q$(C$_{3}$) stops increasing 2 hrs after impact, roughly the same time that $T_{r}$ starts to decrease, however these two profiles together provide a constraint on the chemistry of carbon-bearing molecules during {\em DI}. The excitation profile of C$_{3}$ thus serves as a unique diagnostic of the chemistry and physical processes on short length-scales. 

Future studies could test the effect of variations in the gas density on the excitation temperature by using an analogous molecule, such as C$_2$, to make an independent measurement of the gas density. The 5100\,\AA\ Swan bands of C$_2$ are also recorded in the publicly available Keck spectra, which together with models such as those of \citet{Gredel1989} and \citet{Rouss2001} could be used to quantitatively compare the excitation profiles of both C$_2$ and C$_3$. Such an analysis would also provide a constraint on chemical models of carbon-bearing molecules in comets, since variations in the formation mechanism of C$_{3}$ can change the excitation temperature. Detailed studies of the time-dependent  kinetics of both C$_2$ and C$_3$ will shed light on mechanisms responsible for variations in the excitation profile.

\acknowledgments 

The data presented in this paper were obtained at the W. M. Keck Observatory, which was made possible by the financial support of the W. M. Keck Foundation. The authors wish to recognize and acknowledge the very significant cultural role and reverence that the summit of Mauna Kea has always had with the indigenous Hawaiian community.

\break

\bibliographystyle{apj}
\bibliography{refs}

\begin{table}[h]
\begin{center}
\begin{tabular}{ r c c c }
\hline\hline
Time  &  $T_r$   &  $w$ & $\alpha$  \\  \hline 
8.3    &  48.2 $\pm$   7.4 &	    0.072 $\pm$ 0.015 &   1.02  \\
19.2   &  44.8 $\pm$   8.8 &	    0.066 $\pm$ 0.021 &   0.72  \\
32.6   &  46.2 $\pm$   6.9 &	    0.059 $\pm$ 0.012 &   0.76  \\
48.4   &  47.0 $\pm$   4.9 &	    0.051 $\pm$ 0.006 &   0.88  \\
64.3   &  48.1 $\pm$   4.2 &	    0.054 $\pm$ 0.005 &   0.87  \\
80.2   &  53.7 $\pm$   4.1 &	    0.062 $\pm$ 0.006 &   0.83  \\
96.2   &  60.4 $\pm$   4.5 &	    0.067 $\pm$ 0.006 &   0.84  \\
112.1  &  55.3 $\pm$   3.6 &	    0.065 $\pm$ 0.005 &   0.92  \\
128.0  &  54.0 $\pm$   3.9 &	    0.062 $\pm$ 0.006 &   0.89  \\
143.9  &  50.8 $\pm$   3.8 &	    0.064 $\pm$ 0.006 &   0.81  \\
159.8  &  46.5 $\pm$   3.8 &	    0.063 $\pm$ 0.006 &   0.87  \\
183.2  &  46.9 $\pm$   4.2 &	    0.064 $\pm$ 0.007 &   0.85  \\
214.1  &  45.1 $\pm$   5.1 &        0.065 $\pm$ 0.009 &   0.85  \\
\hline   
\end{tabular}
\caption{Fit parameters for all spectra, which are identified by time from impact to mid-exposure in minutes. Excitation temperatures $T_r$ are presented in K, line widths $w$ in \AA, as well as the unitless scaling parameter,~$\alpha$.}
\end{center}
%\label{tab:myfirsttable}
\end{table}

%% ---Figure-----------------------------------
\begin{figure}
\epsscale{0.90}
\plotone{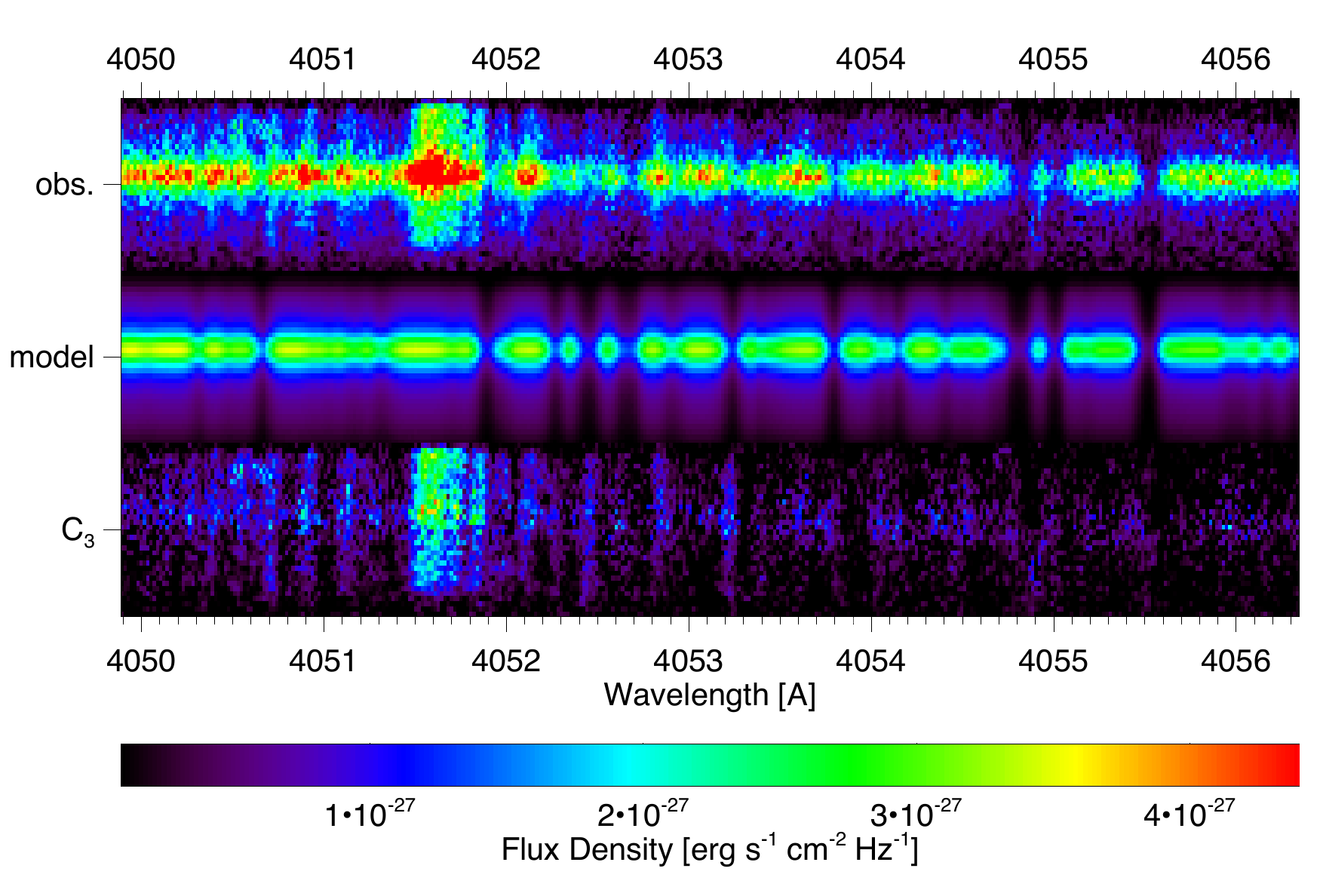}
\caption{An example of the 2-dimensional solar continuum subtraction using the 30 May 2005 UT spectrum of comet 9P/Tempel 1 (obs.). The y-axis is the spatial dimension along the 7'' spectrometer slit. The modeled solar spectrum (see text) is in the middle (model) and the continuum-subtracted 2-d emission spectrum of  C$_3$ is shown at the bottom (C$_3$).}
\end{figure}
%% ---Figure-----------------------------------

%% ---Figure-----------------------------------
\begin{figure}
\epsscale{0.9}
\plotone{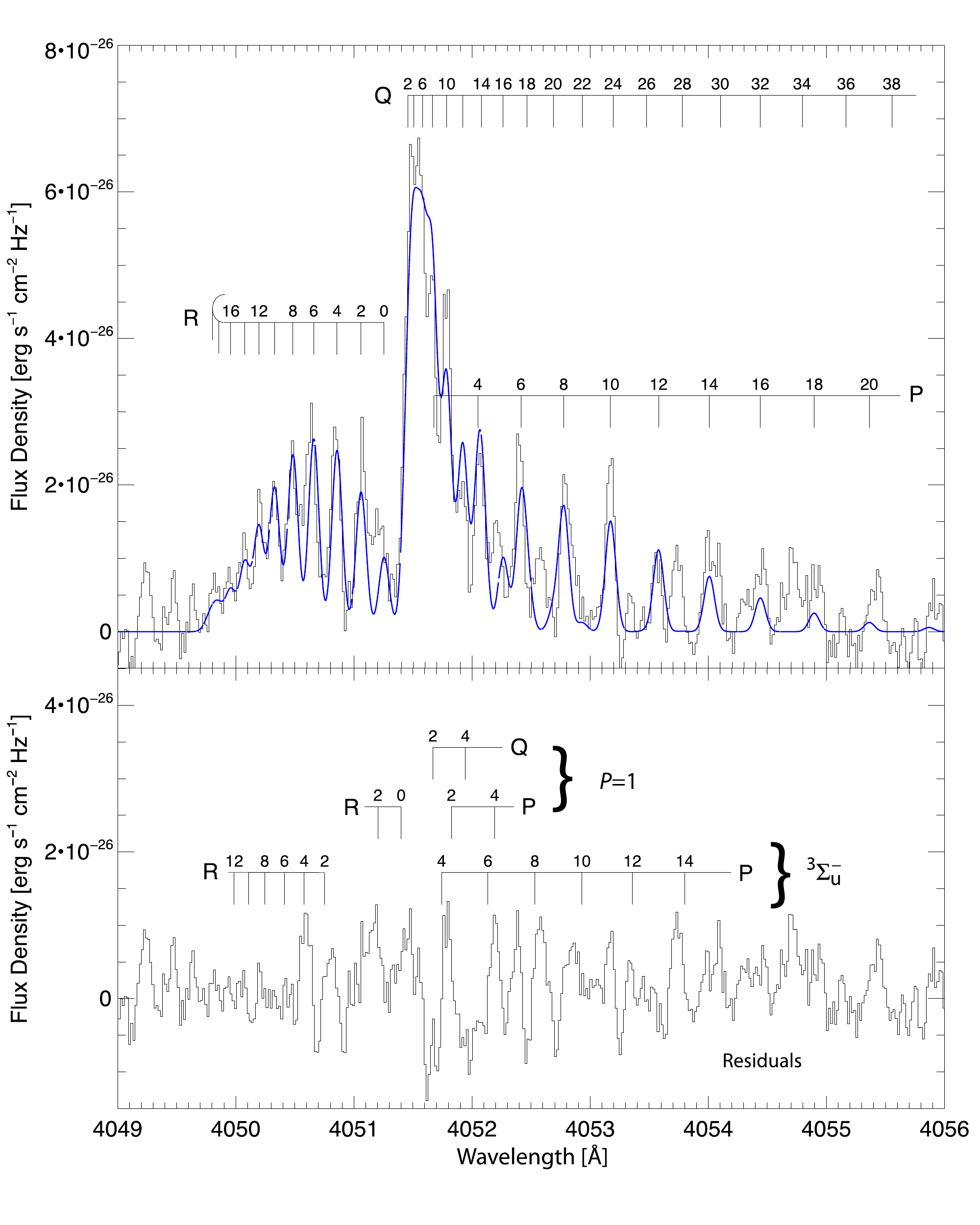}
\caption{The spectrum of C$_{3}$ before {\em DI} (on 30 May 2005 UT) integrated along the slit, with a best fit thermal excitation ($T_{r}=60\pm3$\,K) model (blue) for the rotational population distribution (top). The observed spectrum is the average of three 20\,min exposures. Residuals between the observations and the model, along with laboratory assignments of the transitions from two long-lived perturbing states (bottom).}
\end{figure}
%% ---Figure-----------------------------------

%% ---Figure-----------------------------------
\begin{figure*}
\epsscale{1.0}
\plotone{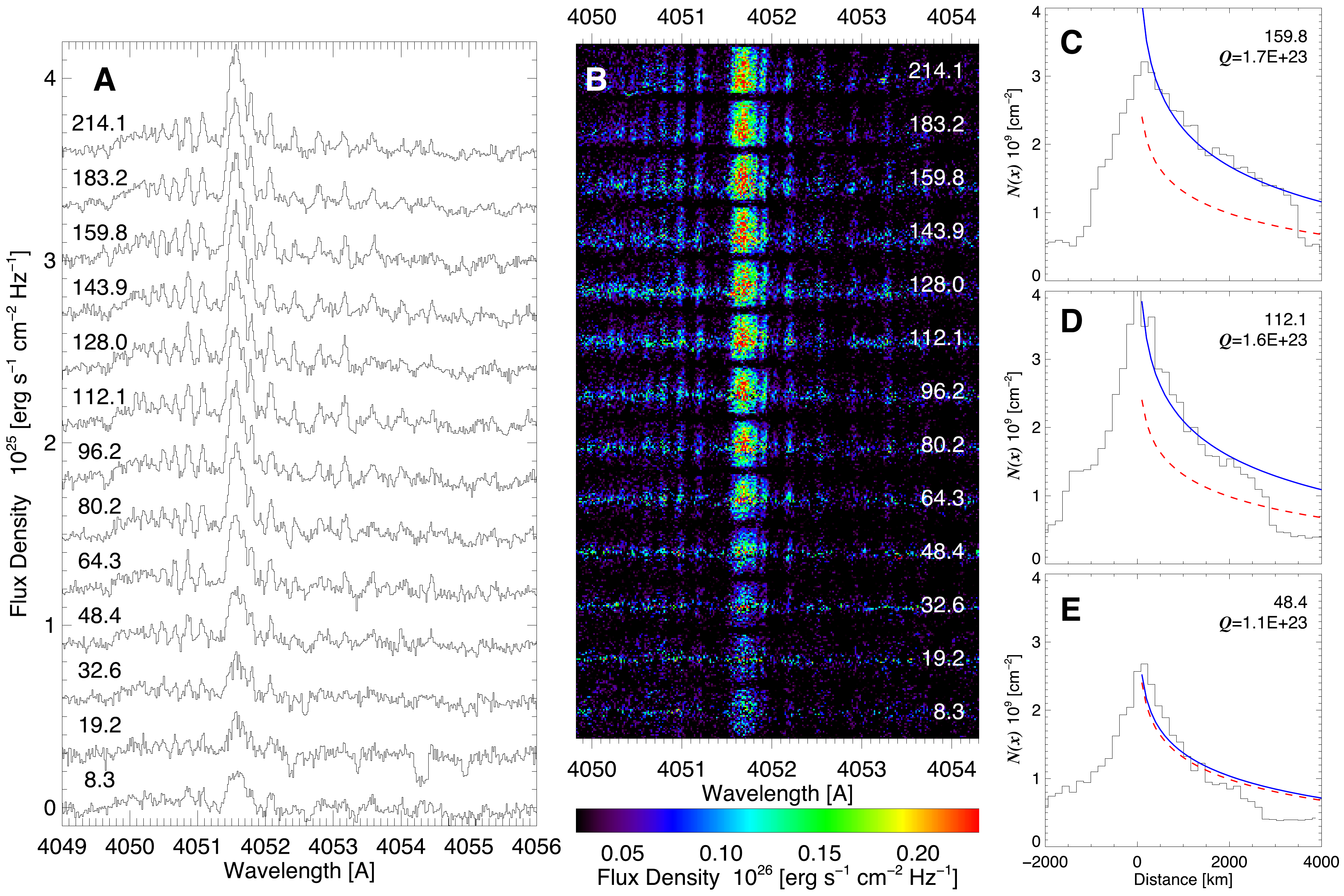}
\caption{(A) Time series of spectra extracted along the slit. Spectra are labeled by the time at mid-exposure from {\em DI} in minutes. (B) The 2-dimensional spectra of C$_3$ after modeling and subtracting the contribution of reflected sunlight off dust. (C)-(E) Examples of the column density of C$_3$, $N(x)$, as a function of the distance from the nucleus ({\em black lines}). For reference, a column density profile calculated using the pre-impact production rate of C$_3$ ($Q$=10$^{23}$\,mol/s) is plotted with a {\em dashed red curve} in each panel. The best fit profile calculated using a Haser model with a variable $Q(\mathrm{C}_3)$ is plotted with a {\em solid blue curve}.}
\end{figure*}
%% ---Figure-----------------------------------

%% ---Figure-----------------------------------
\begin{figure}
\epsscale{0.5}
\plotone{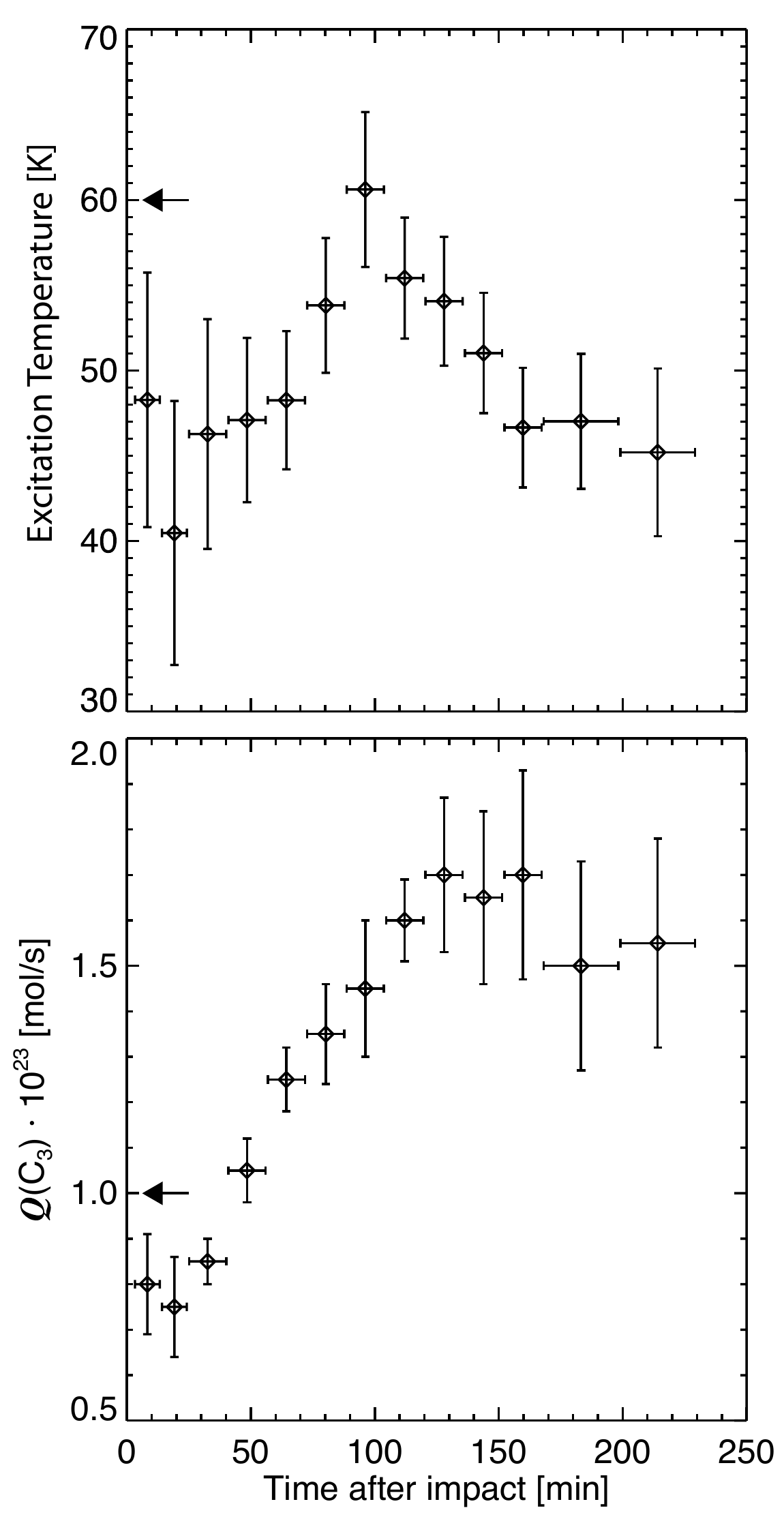}
\caption{(Upper) Comparison of the changes in C$_{3}$ excitation temperature and production rate after impact. $T_{r}$ is measured by fitting a thermal population distribution to the observed spectra and is plotted here as a function of the mid-exposure time after impact. See text for calculation of uncertainties. Error bars along the x-axis indicate the length of the exposure. (Lower) The production rate, $Q(\mathrm{C}_3)$, after {\em DI} determined by fitting the C$_3$ column density profile, $N(x)$, with a Haser model (see Figure~3 for profiles). Left arrows indicate pre-impact values from 30 May 2005 UT.}
\end{figure}
%% ---Figure-----------------------------------

\end{document}